# Names and Affiliation of Authors

1. **Francesca Morselli (corresponding author)**, VU Vrije University Amsterdam
   email: f.morselli@vu.nl
   Adress: Vrije Universiteit Amsterdam. De Boelelaan 1105; 1081HV; HG-0E.
2. **Jetze Touber**, DANS-KNAW
3. **Andrea Scharnhorst**, DANS-KNAW
4.

# Fostering Data Communities - perspective from a Data Archive Service Provider

## Abstract

This paper aims to bridge between the current scientific discourse about the dynamics of data communities in research infrastructures and practical experiences at a data archive which provides services for such data communities. We describe and analyse policies and practices within DANS-KNAW, the Dutch national centre of expertise and repository for research data concerning the interaction with communities in general. We take the case of the emerging DANS *Data Station Life Sciences* to study how a data archive navigates between observation of data research needs and anticipation of research data archival solutions.

This paper offers a unique view of the complex dynamics between data communities (including lay experts) and data service providers. It adds nuances to understanding the emergence of a *data community* and the role of data service providers, both supporting and shaping, in this process.

## Keywords

# Introduction: Data communities in the scientific literature

There is a growing interest in the concept of data communities emerging from the self-organised dynamics of the academic system (Leonelli, 2023). In science studies, science history and philosophy of science, *community* is a widely used concept describing the social organisation of the sciences. The concept of *communities* resonates with other concepts such as *scientific or academic schools* (Amsterdamska, 1987) or *invisible colleges* (De Solla Price & Beaver, 1966), traceable by bibliographic references and collaboration behaviour. There seems to be a preference for the concept of *communities* whenever it concerns social structures in the production of knowledge transgressing the boundaries of *the* academic system, such as *Communities Of Practice* (Lave & Wenger, 1991). One could say that the concept of data communities is another iteration of those related concepts that all try to capture the formation of groups and institutions that share beliefs, needs, and practices.

The term *data communities* has gained momentum in the last decade, attracting the attention of scholars from research fields such as information science, science and technology studies and philosophy of science. However, there is currently no univocal definition of data communities. They can be broadly described as groups of people that create, use or consistently disseminate data for research purposes, administration, or personal interest. An example of a data community is a municipality that collects data about energy consumption to foresee the change in the behaviour of its citizens. Citizens interested in family history, accessing and interpreting data from regional or local archives to trace their family history, can also be described as data communities. Finally, biologists collecting data from farmers about a new crop disease to derive new insights from the data are also a data community. These examples of data communities already point to the variety of their nature. This variety concerns the type of data, the stakeholders involved (municipalities, citizens, researchers) and their possible knowledge and/or interest areas, such as public administration, history or life sciences. The common denominator of these examples is that a data community creates or reuses a specific type of data according to specific methods.

As mentioned above, communities - groups of actors - are widely researched. In research concerning data communities, we find different theoretical and empirical perspectives. Cooper and Springer have focused on defining data communities as characterising the researchers that librarians and archivists support in knowledge institutions (Cooper, D. M., & Springer, R., 2019). Gregory investigated the data communities emerging from specific data uses rather than discipline-specific practices (Gregory et al., 2020). Leonelli and Ankeny have focused on how data communities form around particular methodologies and ways of using and working with data, which they called repertoires (Leonelli & Ankeny, 2015). More recently, Borgman and Groth have studied data communities by focusing on the distance between data creators and data users (Borgman & Groth, 2024). This brief overview shows that the scholarly interest in data communities has grown in the last decade and that research in the social sciences, information studies and philosophy of science is increasingly interested in studying how communities of users interpret and relate to (research) data. The extant literature has also unveiled that data communities are different from scientific communities. Even if research data collections often emerge from empirical practices in one scientific field, Open Science and FAIRification of data practices have stimulated the re-use of data in multiple contexts (Gregory et al., 2020). Also, data practices often connect various scientific communities, which might share data and methods but for diverging purposes. Last but not least, as remarked at the beginning, data collection, curation, and use of data are by no means restricted to the academic system.

Data communities do not emerge in a vacuum. At a closer analysis, data communities interact with organisational, institutional, and infrastructural counterparts, creating and organising the data available for them to access and reuse. Returning to our examples above, municipalities must access data sources from various stakeholders, such as energy providers. Citizens interested in family history interact with institutions such as archives and libraries to obtain access to these data. Finally, biologists gather data about crop diseases by requesting access to farmers' data and local authorities. These examples show that data communities exchange and align the conditions of their existence with organisations and infrastructures, thus creating the boundary conditions for their existence. Our literature research shows that the relationship between institution/infrastructure and data communities has been scarcely investigated. This may be problematic because, without a clear understanding of how these communities interact with the organisations which offer boundary conditions for their existence, we miss a critical understanding of how these data communities grow, evolve, or terminate.

Our paper contributes to closing this knowledge gap by providing empirical evidence from studying the Life Science data community around a specific data service. By analysing the Life Sciences data community and the data archive, which provide its infrastructure and expertise, we will provide insights into how a data community and a data archival service co-evolve. In particular, we will present the example of a life science data community and a research data archive hosted by Data Archiving and Networked Services (DANS-KNAW, henceforth DANS). DANS is an institute of the Royal Academy of Arts and Sciences (KNAW) and the Dutch Research Council (NWO), which offers various research data services. For DANS' archival services, the interaction with the Life Sciences, their multiple communities and their data is a relatively recent development. This makes it a particularly interesting case to study.

This study contributes to the scholarly debate on what data communities are and how they can be defined. By studying an active data community in the life sciences, we provide an interesting case study to the scientific debate, thus supporting researchers in defining data communities and their features. By reflecting scientifically on current practices in an archive, often based on tacit knowledge and experiences, we also contribute to the inner-organisational discourse in establishing shared successful strategies when interacting with data communities. Finally, by studying the features of stakeholders and infrastructures supporting data communities, we identify those elements from which data communities and data service providers can benefit. In the conclusions, we will collect the findings as recommendations for data archives and repositories to improve their services for domain-specific data communities.

The paper is structured as follows. We begin with a methodological section describing the research design and the method to collect and analyse the information. In the "Long-term archiving of digital research data and the role of communities" section, we discuss how DANS, as a provider of archival services, has developed its work with and for communities over the last decades. This excursion into the history of DANS provides context to the Case study section on the DANS Life Sciences Data Station. Concerning our case study, we discuss the blurry boundaries around Life Sciences as a discipline and explain why an institute primarily focussing on social sciences and humanities extended its expertise and services into the Life Sciences. We unravel policy choices and describe DANS' experiences in the interaction with a field, which in itself knows a variety of communities bound by research, data, and/or institutions. We discuss our findings from the case study in the context of this special issue in a Discussion section, "Data communities in practice", and summarise the main lessons learned in the Conclusion section.

# Methodological approach

## Research Questions and Design

In this article, we explore the following research question:
1. How do data communities and data archival services co-evolve?

More specifically, we ask:

2. how does the interaction between communities with data needs and the provision of a service by an institution take place?
3. What is the influence of the data communities on the design of such a data archiving service, and
4. How does the data service create a reference framework for such a community?

We approach these questions based on the understanding that digital research archives are much more than mere providers of services required by specific scientific communities. As unfolded in Borgman (Borgman et al., 2019), digital research archives are mediators between communities' ever-changing data needs and new technological solutions. By providing certain services, they also intervene in those communities.

To achieve these aims, we have set up a qualitative research design to investigate the organisational and relational dynamics of a specific data community (which often is instead a landscape of various communities, as we will see concerning the Life Sciences) with a data archival service at DANS and vice versa. The choice of a qualitative approach lies in the fact that the relational dynamics between a data archive and a data community continuously evolve. Also, in our case, this relationship is in its infancy. As we will see, there are no significant numbers yet, neither in the amount of data to be archived nor in the interaction patterns. A qualitative approach allows us to unveil the intricacies of the interaction between the data archive service provider and various communities with data needs in the Life Sciences. The history of this relationship, including the policy choices along the way, reveals the path dependency and partly also the random character of the emergence of a data community around a service.

Specifically, in this study, we adopt an auto-ethnographic method. Autoethnography, or self-ethnography, is a well-known qualitative research method employed since the 1970s and relies on the description and analysis of personal experiences to understand broader cultural processes (Ellis et al., 2011). Autoethnography has been employed to describe personal and collective experiences, the latter often describing the experiences of minorities or oppressed communities (Pratt, 1991). Our study applies autoethnography to an institutional context, such as DANS and the Life Sciences data communities. Ethnographic approaches in institutional contexts have been employed in fields such as health (Foo et al., 2021) and scientific laboratories (Latour & WooIgar, 1986). In particular, institutional ethnography has studied how people interact with one another in the context of social institutions, such as at school or in the office (Smith, 2005). To our knowledge, there is no proper "way" to report auto-ethnographic methodology. Chang speaks of reporting "styles", such as descriptive-realistic, confessional-emotive, analytical-interpretative, and imaginative-creative (Chang, 2016). We will follow Cooper's and Lilyea's (2022) suggestion to include the following reporting elements: "a review of the literature framing the key concepts; a description of the procedures to collect data; a presentation of the results and a discussion of the findings" (Cooper & Lilyea, 2022).

## Execution of the Design

The study was conducted at DANS, and this object of study was selected for two main reasons: first, all authors are or were based at this institute and have a deep knowledge of the organisation, its technical evolution, its strategy, and its transformations through the years. Secondly, since the end of 2023, DANS hosts the Life Sciences Data Station. Since the evolution of a new data archiving service is one of the leading research subjects of this study, interviewing the manager of this newly available data station has proven particularly insightful.

Cees Hof is the current manager for the Data Station Life Sciences and has been essential in shaping the service by interacting with the research communities for which the service has been designed. He became this article's leading sparring partner and contributed important documentation and text to this paper. As detailed in the acknowledgement, many more colleagues at DANS have been consulted in the process of this research. [Author one] worked at DANS as part of the former DANS Research Group and the current Research Data Expert team and has worked on conceptual models of research infrastructures. [Author two] is currently a senior policy advisor at DANS and was a member of the DANS Management Team for many years, also leading the DANS research group, which executed scientific studies about the institute and its services. [Author three] is one of the two Data Station Social Sciences and Humanities managers responsible for interacting with the humanities. We are aware of the potential bias that may occur since all authors were employed by the institute at some point in time, which now serves as the object of study. Hence, we take special care in describing the methodological approach and the materials used.

## Data Collected

This section introduces the data collected and explains their analysis. We applied qualitative research methodologies: interview, autoethnography (including consultations with other DANS staff), policy document analysis, and further institutional feedback.

For the case study, we conducted two unstructured interviews (Bryman, 2012) between July and October 2024 with the Data Station Life Sciences manager Cees Hof. The preparation of the interviews began a few months in advance with the preparation of a document highlighting the objective and the focus of the interview. We did not prepare an interview protocol because we wanted the interviewee to freely recollect thoughts, link experiences, and narrate his experience about the Life Sciences Data Station. Often, such unstructured interviews are perceived as an extension of participant observation because they are very informed by the environment where they occur (e.g., the DANS office) (Patton, 2002).

The decision to interview Cees Hof, the manager of the Data Station Life Sciences, emerged at the very beginning of the brainstorming phase of this study. He joined DANS in 2016 to seize the data opportunities within the Life Sciences, and since 2022, he has officially become the DANS Data Station Manager of Life Sciences. Cees graduated as an aquatic ecologist and worked as a PhD and postdoctoral researcher in evolutionary (paleo)biology. His expertise in research data developed when coordinating European projects around biodiversity data and specifically when managing the Dutch Node of the Global Biodiversity Information Facility (GBIF) for more than 10 years. His knowledge of the different research domains within the Life Sciences has been pivotal when expanding DANS' data archiving service towards this field of research and probing the data opportunities relevant to DANS. The interviews with Cees Hof were recorded and consequently transcribed for further analysis.

Besides the conducted interviews, we also based this study on several informal consultations. While this data cannot be quantified due to its ephemeral character, it has nonetheless been essential in building awareness about the importance of studying the life science data communities at DANS. The authors of this paper, together with Cees Hof, met regularly once a week over several months. The content of these informal chats has been annotated on personal block notes and later reviewed as research data for this study. However, it is important to note that these notes precede the beginning of this study, so they do not follow the rigour of participant observations during fieldwork (Marshall & Rossman, 1989).

Apart from the interviews and the informal exchanges, we also analysed strategic policy documents produced by DANS before creating the Data Stations. Among them are DANS Strategy documents formulated from 2014 to 2024, other reports on the archival services, guidelines about the services, and earlier scientific studies about the archive. The *Beleidsnota DANS en de Levenwetenschappen* (policy document on DANS and the Life Sciences) had an important role in the case study, as it details the central motivations to include the life sciences in DANS' data archive (Hof, 2019). The analysis of these documents has been crucial to familiarise us with the institutional and strategic aspects of DANS, as it showed us the reasons and the method of including the life sciences in DANS' network and the user base of its archival services. By alternating between strategy policies and other documents from the past, on the one hand, and consultations, on the other, we could reflect on DANS' culture when it comes to serving and interacting with communities around the archival services. In this paper, we re-emphasise the active mediating role of the archive. We believe that to understand the nature of data communities, it is important to look at their interactions with the institutions they use the services. Consequently, in the following section, we present some aspects of the DANS institutional history with a special emphasis on the interaction with communities: research communities, so-called designated communities and data communities.

# Long-term archiving of digital research data and the role of communities

## The foundation of DANS as an institution and its data services

The history of DANS mirrors the growth of digital research data over the last two decades and its increasing importance for current scientific research. The roots of this development date back to the start of the automation of research data in the pre-Internet era. It is helpful to first unfold DANS's institutional history to understand how deeply the sense of serving specific research communities is encoded in its cultural DNA.

DANS was founded in 2005 as a spin-off of the *Nederlands Instituut voor Wetenschappelijke Informatiediensten* (NIWI - Netherlands Institute for Scientific Information Services) built around the former library of the Royal Netherlands Academy of Arts and Sciences. Three parts of NIWI later became cornerstones of DANS: the Steinmetz Archive, the Dutch Historical Data Archive (NHDA) and the Scientific Statistical Agency (WSA). The Steinmetz Archive was founded in the 1960s to preserve social science quantitative data in a machine-readable format (e.g., magnetic tapes, punched cards). The archive was first stored at the University of Amsterdam and then merged with the KNAW collection in 1972 (GESIS, 1989). The University of Leiden first hosted the Dutch Historical Data Archive (NHDA) before joining NIWI (Blauw, 2005). Finally, the Scientific Statistical Agency (WSA) was an intermediary organisation that supported researchers in retrieving research datasets from the Centraal Bureau voor de Statistiek (Statistics Netherlands). At the beginning of DANS, one early project, eDNA, created the cornerstone for the specific role of DANS for the Dutch archaeological community (Gilissen, 2014), which profoundly influenced DANS collaboration with communities.

Up to the early nineties, retrieving datasets was often regulated inside scientific communities and was an expensive and challenging endeavour (Borgman et al., 2015). With the introduction of computational methods in all scientific fields, digital research data and retrospective digitisation of research data were introduced. In the Netherlands, DANS pioneered a web-based service called EASY (Electronic Self Archiving System), the first version of which went online in 2007. This service paralleled other data-sharing services, such as the public Dataverse Network, which started as a service in 2006 (King, 2007), the commercial open access repository *FigShare*, and the now widely used CERN-based service *Zenodo*, which was launched in 2013. By the early 2000s, DANS had a data archive in place, which allowed individual researchers and members of specific research communities to deposit datasets for archiving and publication.

At DANS at this time, one would not use the concept of data communities so much but instead refer to the concept of so-called designated communities, which partly entails data communities. This term comes from the conceptual framework under which DANS data services operate, namely the Reference Model for an Open Archival Information System (OAIS)[1], an ISO standard widely used among service providers for archiving data. This standard defines three different types of actors in a service provider's environment: producer, management, and consumer. In the consumer category, a 'designated community' is introduced as a specific class next to other types of consumers. Its definition reads: "Designated Community: An identified group of potential Consumers who should be able to understand a particular set of information. The Designated Community may be composed of multiple user communities. The Archive defines a Designated Community, and this definition may change over time." (CCDSD 2012, p.1-11)

To become aware of the subtle differences between the terms *designated*, *data* or *research community*, it is interesting to notice that Lavoie (Lavoie, 2004), in his introduction to the OAIS standard, uses the term 'community' in various ways: data community from the needs of which this standard has developed (for the establishment of OAIS that was primarily the community using space data); then the designated community (as a specific concept in the OAIS model defined above), and with the notion of a user community of an archive (which can change in time).

---

[1] "Early draft versions of the OAIS reference model are available at http://nssdc.gsfc.nasa.gov/nost/isoas/ref_model.htm. The first approved version of OAIS (CCSDS 650.0-B-1 [Blue Book], equivalent to ISO 14721:2003) is also available at this link. The current version of OAIS is available at http://public.ccsds.org/publications/archive/650x0m2.pdf (CCSDS 650.0- M-2 [Magenta Book]) or at http://www.iso.org/iso/iso_catalogue/catalogue_tc/catalogue_detail.htm?csnumber=57284 (ISO 14721:2012)" (Lavoie, 2012)

Traditionally, archives evolved around specific bodies or institutions ((Ketelaar, 1996; Cook, 2011). However, with the emergence of the Internet and networked information services, which can be globally used, it is no longer a specific institution representing a community for which the archival service is designed. So, the definition of the designated or user communities requires a reinspection.

As we talk about research communities as one of the possible user communities, inevitably, classification schemes to identify research communities, disciplines, and subfields become important. The Preservation Plan for EASY ("Preservation Plan for EASY," 2017) details the disciplines for which the data service is provided:

"The Archive's holdings consist of:
- Data from the **humanities** include texts, spreadsheets, databases, images, transcripts (including time-based transcripts), audio and video files, and, more recently, graph representations of knowledge (linked data formats).
- Data from the **social sciences**: mainly quantitative (statistical) data, questionnaires, codebooks, test responses and some qualitative datasets, which include field notebooks, interviews and interview transcripts.
- **Archaeological data**: reports, data from excavations and trial trenching and post-excavation analysis, such as texts, databases, spreadsheets, GIS files and images (vector and raster images).
- Data from **geospatial sciences**: vector graphics, CAD drawings, GIS files.
- Data from the **life sciences and health research**: images, statistical data, spreadsheets, databases, laboratory notebooks and texts.
- Across those disciplines and beyond: Data in Linked Open Data formats, models, algorithms, scripts, executables." (Section 2.2 Characterisation of the content)

Interestingly enough, in 2017, the Life Sciences, next to Health Research, was mentioned. We come back to this in the Case Study Section.

The Preservation Plan document also defines the interaction with users. "The designated community of the Archive consists predominantly of humanities and social sciences scholars. The Archive monitors the community through substantial contacts, for instance, during data acquisition and ingest, in applied research projects, membership of European Research Infrastructures, pilot studies with data producers, via training & consultancy and by offering discipline-specific services." (Section 2.1 Mission) ("Preservation Plan for EASY," 2017).

Each data set deposited in EASY was annotated using a metadata scheme based on the international Dublin Core™[2]. In this scheme, one field (the subject field) indicates the topic of the dataset, as is usual in bibliographic indexing. In EASY, DANS applied the NARCIS classification tree for subject indexing (initially developed for the Dutch Research Information system (Dijk et al., 2006; Braukmann, 2020).

A couple of visualisations of the content of EASY were produced around 2012. Figure 1 shows the distribution of datasets across subjects or specific research fields. The size of the circles drawn around specific nodes in the classification tree of research fields indicates how many datasets in the collection (at the time of 2012) have been tagged with those research field categories.
One immediately sees the specific role of archaeology, which has since constituted the most extensive collection inside the archive. However, one can also see the large variety of fields in the social sciences and humanities that were represented in the data archive in the early days.

[Figure 1]

To summarise, EASY took "a universal approach" - a one-size-fits-all data archive - where social sciences and humanities communities seem to be well served by a single system, avoiding data silos with boundaries defined by disciplines.

Let us briefly return to the specific place of the archaeological collection in EASY. The significant presence of archaeological data can be explained by the DANS archival service's specific role in the Dutch field of archaeology. In the Netherlands, according to the guidelines of the Rijkserfgoeddienst and the Faculty for Archeology at Leiden University, archaeologists must deposit their data in the E-Depot for Dutch Archaeology, EDNA, which was provided by DANS, and became part of EASY. Several new technological solutions were implemented around the archaeological collection, partly funded by participation in European projects. Among them are additional field-specific controlled vocabularies and an interface to search for datasets by geographic location. Moreover, the domain-specific information exchange protocol and an automatic uploading pipeline have been established for an efficient submission process. The size and ongoing growth of archaeological data can be attributed to institutional settings and regulations rather than a large archaeological data community, even if the fact that Leiden University has its faculty for Archeology certainly contributed to this success. However, as the archaeological case shows, this success also comes with accommodating specific needs, which a one-size-fits-all approach can not serve.

---

[2] https://www.dublincore.org/specifications/dublin-core/dcmi-terms/

The case of archaeology, which we only briefly sketch here, already highlights elements that we will later discuss for the case of the Life Sciences, namely: collaboration with institutions, the role of legal frameworks, the role of scientific and other communities (as in archaeology, there is also a wider group of citizen scientists), and new technological opportunities.

The success of archaeology - growing to have the largest share of the datasets deposited in EASY - also triggered further discussions in DANS on how to best accommodate the growing and changing wishes of various communities, institutions, and research infrastructures. Both the metrics indicative of the nature of data deposits in the EASY archive and the archival policies as expounded in the Preservation plan, thus highlight the disciplines of Social Sciences and Humanities, including Archaeology, as the principal beneficiaries of the DANS repository. This already pointed towards the subsequent evolution towards a community-oriented approach rather than an individual researcher-oriented approach in further developing the archival services. At the same time, the relevance of other communities, such as those in the life sciences, was also already within sight.

As we will expound in the following subsection, this eventually led to the current Data Station Model. In the following subsection, we discuss how the increasing awareness of the important role of research data from the science policy side (as well as the ongoing progress in the digitisation of sources and digital methods and the formation of communities around data) informed processes in DANS to innovate its services and to give communities an even more prominent role.

## The DANS Data Stations - explicit community orientation

In the previous section, we described the early history of the archival service of DANS and the role communities played in designing, implementing and executing this service. Between 2005 and 2022, the data collection held by EASY grew substantially. The latest systematic analysis stems from 2019 (Doorn, 2020) and shows a substantial growth in the number of datasets and their size. The last self-evaluation of DANS lists more than 250k datasets (*DANS Self-Evaluation-Report*, 2024) compared to the 20+k datasets about 10 years prior. This growth was triggered by further digitisation of sources in many scientific fields (e.g., in the social sciences and archaeology) and by digital research methods becoming the norm (e.g., data and code being released alongside journal articles). At the same time, the field of science policy began to refer to data as the 'new oil' of the sciences (Borgman, 2010; Γληνός & Glinos, 2010; Christine L. Borgman, 2015). In addition, the emergence of the Open Science movement fostered the awareness that archiving research data in bespoke data services was an essential component of sound research practices (Wilkinson et al., 2016).

When EASY started twenty years ago, DANS was still a pioneer in digital (self-) archiving. By now, there are countless projects funded at the interface of research and research infrastructures, including the European Research Infrastructure Consortium (ERICs), which develops policies and services around data, the European Open Science Cloud (EOSC), and globally, organisations such as the Research Data Alliance (RDA) which co-create and operate in an emerging Research Data Landscape. In other words, the past two decades have witnessed such a growth in the volume of data produced and such a multiplication of stakeholders that reorganisation of labour behind the services provided and a division of labour was inevitable. DANS discussed intensively how, as an organisation, it could adapt to such changes.

One thread of discussions concerned the emergence of more stakeholders around research data. Between 2015 and 2020, the increasing number of institutional repositories at Dutch Universities (e.g., the 4TU repository) animated a discussion inside DANS on dividing tasks and responsibilities among institutions providing support for specific data communities. For this reason, DANS introduced the so-called *Front-Office-Back-Office* model (Doorn et al., 2014). The main idea was that university libraries, national data services (e.g., DANS) and basic technical e-infrastructure (e.g., SURF) would have complementary responsibilities and could work together in a federated infrastructure. In this vision, DANS services were initially thought to be better placed in a 'back-office' focused on long-term preservation, archiving, and other aspects of shared data services. Following such a model, contact with the designated communities or data communities (data producers and consumers) would be delegated to the institutions supporting a 'front-office' function, such as university libraries and domain-specific research infrastructures. The graph below represents the Dutch federated data infrastructure reported in an internal strategic document (Doorn et al., 2014).

**[Figure Two]**

Around the mid-2010s, DANS discussed how it could serve data communities not yet supported by domain-specific institutions (e.g., universities) and large research infrastructures operating at the national level (e.g., local branches of the ERICs). In these discussions, the concept of long-tail research data became important. "Long tail data" are produced by small teams or individuals, are usually small in volume, heterogeneous and are often local in the sense that they are created from place-specific research projects (Wallis et al., 2013). Moreover, while "big science" and "big data" are better provided for institutionally and financially, "small science" and "long tail data" constitute the majority of the available data in the research landscape (Heidorn, 2008; Wallis et al., 2013). They also come with specific challenges as the related communities are small and often not yet mature technology-wise. Also, long-tail data often are situated at interfaces between various scientific fields. For all these reasons, long-tail research data has difficulties being archived by universities or other official repositories. Therefore, the research communities creating these data could represent a promising sparring partner for the further evolution of the DANS data archive services and its expertise around research data in general. One could say that DANS moved its services from 'individual pioneers' in research data production as targeted by the self-archiving service of EASY to 'pioneering data communities', which also come with non-standard (or not yet standardised) needs.

At the beginning of the 2020s, making long-tail data discoverable and reusable became thus DANS' primary focus, especially by serving data communities that are often small and scattered and that come with specific data in small volumes, for which a research data infrastructure is not yet set up (Doorn, 2020).

For this reason, in 2020, DANS began planning the Data Stations as community-oriented data archiving services (Doorn, 2020; *Focus on FAIR*, 2020). The strategic document (Focus on FAIR) introducing the data stations mentions that: "DANS will offer a combination of 'reuse' and 'long-term preservation' services. The technological foundation of the DANS infrastructure will be the Dataverse software, developed and maintained by an international community coordinated by Harvard University (https://dataverse.org/). Multiple instances of Dataverse – referred to as Data Stations – will be customised to the needs of various disciplines and research communities. Researchers can bring their data to one of the DANS Data Stations: Archaeology, Social Sciences and Humanities, Life Sciences, and Physical and Technical Sciences. Each station is assigned a Data Station Manager to establish and maintain contacts with the relevant research community." (*Focus on FAIR*, 2020)

This summary of the Data Station approach combined several essential changes: a change of the technological base (i.e., from EASY (based on Fedora) to Dataverse), the establishment of community-oriented archives, and the introduction of Data Station managers who also operate as data community managers. Finally, in a meticulous and time-consuming process, the EASY content migrated to the Dataverse-based Data Stations, which went live consecutively (Data Station for Archeology in 2022, Data

Station for SSH in the spring of 2023, Data Station Life Sciences and the Data Station Physical and Technical Science at the end of 2023).

In the following section, we will unfold how archival services and data communities co-evolve around the establishment of one of the newer Data Stations, the Data Station Life Sciences. This will also clarify how the data stations have been designed to serve data communities connected to long-tail research data.

# Case Study: The Life Science Data Station and the Long Tail of Research Data

Following up on the reflection of the previous sections about the relationship between an archival service provider and data communities, we will now zoom in on the relationship between data communities for the life sciences and the DANS Data Station Life Sciences, and address the research question: how do a data community and a data archival service co-evolve?

This section has two parts: first, we will discuss life sciences as a scientific field and its specific data needs; second, we will present the DANS Data Station for Life Sciences. Concerning the Life Sciences in general, we begin with a sketch of the location of the Life Sciences on the large-scale maps of science (Borner, 2010). We complement this by describing the current Dutch institutional landscape around the Life Sciences, the research data information needs and the maturity of data practices in the field. We continue describing the original plan for the Life Science Data Station. The second part, dedicated to the DANS Data Station for the Live Sciences, starts with describing the current content of the Life Science Data Station collection. We progress by unfolding the various channels and interactions with stakeholders and related data communities in Life Sciences in the Netherlands.

# The Life Sciences as a discipline, its institutional representations in the Netherlands and its data practices

## The Life Sciences as a discipline

The naming and classification of scientific disciplines is always debated (Weisz, 1982). According to Wikipedia, the Life Sciences encompass all branches of the (natural) sciences that study *life* (from microorganisms to human beings) ("Life Sciences," 2024). The Life Sciences is the branch of the Natural Sciences that studies the 'living', while the Physical Sciences study 'non-living' matter. In the Universal Decimal Classification, one of the oldest modern generic knowledge classifications (Rayward, 1991; Slavic, 2005), class 5, encompasses *Mathematics, Natural Sciences,* under which we find *Environmental Sciences, Biological Sciences, Botany and Zoology*, but not the term Life Sciences. The Field of Science (FOS) classification used for OECD statistics on research and development does not mention the term *Life Sciences* either. Its main fields are Natural Sciences, Engineering and Technology; Medical and Health Sciences; Agricultural Sciences; Social Sciences and Humanities. However, NARCIS, the original Dutch classification system for research information (Dijk et al., 2006), lists Life Sciences as a concept under Life Science, Medicine and Health care.

One of the most thought-provoking maps of science is the so-called University of California San Diego (UCSD) (*Map of Science - UCSD Map*, 2007), which visualises networks of scientific communication as connected entities of a 'globe of knowledge'. Also, in this map, we do not find the term *Life Sciences* but *Biology,* next to *Biotechnology* and fields of the medical sciences.

**[Figure three]**

In conclusion, what could be summarised as Life Sciences forms a bridge between the natural sciences (those dealing with the non-living, such as physics and chemistry), medical sciences and health care. (Börner et al., 2012).

This overview of the classification of scientific disciplines shows that life sciences is a heterogeneous area, with many fields under its umbrella. This also leads to many different data types, data needs, and related data communities.

## The Dutch Institutional Landscape in the Life Sciences

In the Netherlands, as in many other countries, the Life Sciences research landscape is large and complex, hosting various institutes and organisations. The scale ranges from large University Medical Centres (UMCs), hosting hundreds of specialist staff members, PhDs, and postdocs involved in research, to small start-ups and individual freelance researchers. Academic, non-profit, and commercial research areas range from traditional ecological field research to advanced medical biotechnology.

To provide a rough impression of institutes and organisations: of the currently 14 public universities in the Netherlands, including the 4 Technical Universities, 12 have faculties or large research groups linked to the life sciences. There are seven large University Medical Centres in The Netherlands, one (Amsterdam) with two major localities (*About the Umcs | NFU*, 2021). 27 Smaller Dutch hospitals have joined forces with the Foundation Collaborating Top Clinical Hospitals (STZ) to promote research as one of their primary goals (*STZ*, 2024). Of a total of 21 academic research institutes funded by the Dutch Research Council (NWO) and the Royal Netherlands Academy of Arts and Science (KNAW), four are fully dedicated to Life Science research: the Dutch Institute for Ecology (NIOO); the Westerdijk Fungal Biodiversity Institute; the Hubrecht Institute, the Dutch Institute for Neuroscience (NIN), and one partly: the Royal Netherlands Institute for Sea Research (NIOZ). Finally, at least half of the 36 Dutch state-funded Universities of Applied Sciences perform research and offer curricula in Life Science-related disciplines (such as health care, agriculture, veterinary sciences, environmental sciences, green educational sciences).

Several prominent governmental knowledge institutes (in Dutch, *Rijkskennisinstellingen*) and institutes for applied research (such as the Dutch National Institute for Public Health and the Environment - RIVM - and Naturalis Biodiversity Center) host departments for researching applied and fundamental Life Science. Concerning the monitoring of health and environmental conditions and the associated research, the Netherlands hosts numerous networks at different geographical and/or administrative levels, such as 25 Public Health Services (GGDs) at the council level, or the 21 historic Water Boards (in Dutch *Waterschappen*) that examine, amongst other characteristics, the biological quality and biodiversity of surface waters. These figures are reflected in the relatively large share of publications, especially the medical sciences, when comparing Dutch scientific publication output to other countries (Rathenau Instituut, 2024).

Concerning research data created by the Life Sciences, the landscape is even less clear. Data originates from research at all biological organisational levels (Eronen & Brooks, 2024), from research on atoms and molecules to studies of the Earth's biosphere. Advanced technological developments of the last decades, such as the Polymerase Chain Reaction (PCR) for identifying gene sequences or bioimaging technologies like Magnetic Resonance Imaging (MRI) and Cryogenic Electron Microscopy (cryo-EM), have led to a steep increase in data volumes. In addition, less techno-driven research, such as long-term cohort studies or citizen science field research, has expanded. However, it is difficult to trace who is creating what data and in what quantities. The same applies to used data standards, formats, data storage systems and exchange protocols.

## The current content of the DANS Data Station Life Sciences

The DANS Data Station Life Sciences currently hosts data from health and medical science and data from the "green" Life Sciences such as Earth and Environmental studies or ecological and agricultural research. It is a relatively small collection (around 1000 datasets) compared to the Data Station Social Sciences and Humanities or other institutional repositories (e.g., 4TU). As Cees Hof points out, it is a relatively young collection with a shorter history than the other data stations. The lack of direct links with the Life Science communities during the early days of DANS is reflected in the few datasets deposited in these early days because they are published as provided, without domain-specific enrichments. On the one hand, the internationally accessible nature of the previous DANS-EASY repository has generated a rather diverse collection of datasets. On the other hand, the Data Station holds unique Dutch research material which broadly represent Dutch Life Science research institutes and disciplines.

At the beginning of the Data Station Life Sciences, the strategic focus was on catering to the aforementioned "long tail" research data. During the interviews and consultations, Cees Hof referred explicitly to the distinction between the "big science data" and the "long tail of research data". We introduced this distinction earlier, which has been discussed in the literature too.

Within DANS, the concept of the long tail research data has been used in several communications and publications, such as the PLAN-E workshop and report in 2018 (PLAN-E, 2018) and the Workshop for FAIR Data for the "Long Tail of Science" (Lorentz Center, 2021). The DANS Data Station Life Sciences aims to help the long tail of research data to be findable and reusable. The assumption is that by using domain-specific (meta)data standards, the long tail of research data can become part of big science. For example, suppose somebody (such as a citizen or a researcher) has a small dataset on butterflies from the Veluwe National Park. In that case, those data can be stored and made accessible in different formats, including data standardised in a format for a specific use, like biodiversity research. If those specific data can be made discoverable and harvestable by machines, then they can quickly become part of Big Science efforts, such as a project creating a Digital Twin of the ecosystem of the Veluwe. By providing the technical means within the data station, we help researchers turn the long tail of science data into big science data.

Other datasets the Data Station Life Sciences aims for are typically generated by a diverse team of researchers from different institutes and organisations and where a non-institutional repository is preferred. As DANS is not directly connected to a research institute and has the sole purpose of preserving and sharing data optimally, it operates independently from data policies established at universities and research institutes. For this reason, it offers equal conditions and credits to the data owners, regardless of their institutional origin. The Data Station Life Sciences also aims to act as a "project repository" or "community repository" for Dutch research infrastructures and distributed research communities searching for a central place to archive their data. In these situations, DANS acts as a partner that can adapt its service to the needs of the infrastructure and/or the community. The way the DANS Data Station Archaeology operates with the Dutch Cultural Heritage Agency (RCE) and the e-depot for the Dutch Archeological community is a good example of such a mutually beneficial relationship.

## The role of data communities and mediating network structures in shaping the Data Station

The DANS Data Station Life Sciences is also a good example of the co-evolution of new research infrastructures and specific data services. From the beginning, this Data Station has profiled itself as a data station for the health, medical, and green life sciences, with three "subdomains" also distinguished in the Dutch National Roadmap for Large-scale Research Infrastructure. Partly because of the experience of the Data Station manager with the "green" Life Sciences (e.g. ecology, biodiversity and agricultural studies) and because of his past working relationship with the Netherlands Institute of Ecology (NIOO), the first project-based collaboration was established with LTER-LIFE, an ecological Research Infrastructure coordinated by the NIOO (*LTER-LIFE*, 2022). Within this long-term infrastructural project related to the international Long Term Ecological Research (LTER) Network, Digital Twins of ecosystems are being created to allow research in a strictly digital environment. Scientific data is key to creating Digital Twins (virtual replicas of physical systems). LTER-LIFE focuses on two well-studied ecosystems in the Netherlands, the Wadden Sea and the Veluwe. Data for and from LTER-LIFE should be Findable, Accessible, Interoperable and Reusable (FAIR) too, and for this reason the DANS Data Station for the Life Sciences has been involved. The data archiving requirements of the LTER-LIFE community have been discussed in advance and captured in the project descriptions. In turn, the Data Station manager provided, and still provides, directions based on DANS's longstanding experience with data repository services.

As noted above, a substantial part of the Dutch life sciences research community focuses on health and medical research. Discovering and developing the proper niche within this landscape has been challenging for the DANS Data Station Life Sciences. A fortunate development within this context is the establishment of a National Health Data Infrastructure in the Netherlands, *Health-RI*, that started in 2021 after receiving a multi-million investment from the Dutch Government. Within Health-RI, a FAIR health data architecture is one of the main deliverables for which DANS is responsible. Also, within this development, the initial focus of DANS has been on how to make the "long Tail of Science" datasets (that the Data Station Life Science hosts) part of the Health-RI data infrastructure. Collaboration and the exchange of experiences within HEALTH-RI currently focus on technical issues like the interoperability of metadata standards and the use of domain-specific knowledge Organisation Systems (SKOS). Furthermore, the DANS Data Stations Life Sciences and SSH are involved in the European *QUANTUM* project, developing a quality label for health data hosted by the European Health Data Space (QUANTUM, 2023). Health-RI is a major participant in the QUANTUM project, and through this European collaboration, the Health-RI - DANS collaboration is also strengthened.

# Data communities in practice - Discussion

## Scientific implications

This article studies the Life Sciences Data Station at DANS-KNAW as an example of how existing and emerging data communities in the life sciences interact with a data archive that hosts research data for and from their communities.

This study is among the first to analyse the dynamics between a data community and a data archive. The perspective provided by this article represents a novelty in the study of data communities. Previous reflections on this theme by scholars such as Borgman, Gregory and Leonelli have inspired this article. In particular, from Borgman, we draw the notion of data communities as knowledge and information infrastructures. We were inspired by Gregory's reflection on specific data use and reuse practices, overcoming the traditional disciplinary differences. Finally, from Leonelli and Ankeny, we draw on the notion of repertoires, a concept describing a specific set of practices that actors put in place to describe, connect, search, use and share research data. While these authors have significantly contributed to describing the dynamics and features of data communities, they do not yet examine in detail the institutional and organisational context in which data archives and data communities negotiate their existence.

The comprehensive approach presented in this article enriches the reflection on data communities, showing that data communities exist in complex realities where not only disciplines, data formats and use/ reuse practice are relevant: institutional settings (e.g., data generated in small local projects or by large research infrastructures) are crucially important in determining the life of a data community and where these data will be hosted. We presented the case of data generated by small projects and small institutions (what has been called 'the long tail of research data"), and the difficulty of finding an appropriate host for their data. This is illustrated in the above mentioned example about the butterflies datasets from the Veluwe

National Park: in this case the absence of an institution able to host this dataset would limit its accessibility. In such cases, data archives such as provided by DANS and its Data Stations, are valuable partners to further design these data communities' future.

In addition, this article contributes to the discussion on the definition of a data archive, especially of a Data Station. In the past decades, many universities, following national and international open science guidelines, have implemented their repositories, created digital infrastructures and set up support services (e.g., data stewards) for researchers. Consequently, data archives like DANS, pre-existing university repositories, had to redefine their missions and objectives and cater to research communities outside the traditional avenues of knowledge creation (e.g., universities). We described this development and provided evidence for DANS modifying its own mission. As a consequence of the changing data landscape in the Netherlands, the Data Stations allowed DANS to tailor their offer to specific data communities by offering them a repository alongside documentation, information (e.g., about data formats) and training. In conclusion, the static image of a data archive does not fit DANS-KNAW, which has become a translator, a connecting point, between data communities, archives and technology (*DANS Self-Evaluation-Report*, 2024).

## Practical implications

In addition to the aforementioned scientific implications, this article's reflections have practical implications for data archives, too. For example, data repository coordinators and managers might benefit from reading reflections as this article provides because it makes a novel perspective on data archives explicit: it shows the role of data archives as active stakeholders in the research data landscape, one where their expertise is offered as consultancy to research communities and projects.

Furthermore, it portrays the evolution of an existing data archive, highlighting the strategies implemented to adapt to the changed landscape of research data described in the previous section. This understanding of an evolving archive, both adapting and guiding the needs of the data communities, is helpful for repository managers and coordinators to reflect on their jobs as continuously evolving, too.

Secondly, data archives managers and coordinators might benefit from this article because it highlights the entanglement of technological development (i.e., the technology of the repository) and the way the relationships with data communities evolve. We have seen that the explicit community orientation leads to another technological response to specific community needs in metadata schemes, data formats, and interaction interfaces. This aligns with the current discourse on implementing Open Science and FAIRification principles in various communities and with various service providers, all contributing to a federated Europe-wide data research infrastructure.

# Conclusions

This study reflects on the evolution of a data community (specifically, in the Life Sciences) and a data archive. We interviewed the manager of the Data Station for the Life Sciences, Cees Hof, and extended our investigation to the analysis of strategic documents and informal exchanges with various experts within DANS. In this study, we provided a case study showing a data community and the data archive hosting its research data. What emerged from our reflection is a data community and a data archive evolving, reciprocally creating the conditions for their existence by following their changing needs and features.  The novelty of this approach lies in examining both data communities and data archives, both individually and in relation to each other. The institutional aspect of data communities and the archives hosting their data are crucial of this relationship.

Such a vision of evolving data communities and archives entails a novel perspective and requires a transition toward responsive and reflexive strategic approaches. Responsive and Reflexive approaches have gained the attention of social scientists and evaluation experts in the past few years for mobilising the emergence of social, political, and ethical considerations in the "technocratic process of producing numbers." (Bandola-Gill et al., 2023). For a data archive, being responsive means establishing a connection with the data communities and anticipating their evolution; on the other hand, a data archive should not only be seen as merely following the needs of data communities but as an innovator, too, guiding the data communities through changing technologies, data formats and open science best practices.

In conclusion, this study contributes to bridging the gap identified in the opening of this article on the relationship between data archives and data communities. We provided the example of a data archive and a newly formed data community within the life sciences and followed their evolution. We reflected on the socio-technical dynamics that took place between the two stakeholders. Despite portraying a single case, similar dynamics occur for other data communities and data archives because we reflect on topics (such as technology, data requirements, and formation of data communities) that can be extended to other case studies.

# Acknowledgement

A special acknowledgement goes to Cees H.J. Hof, the Data Station manager for the Life Sciences Data Station at DANS. Cees was available for interviews and regular consultation, provided internal science policy documents and actively contributed to the text, in particular to the Case Study Section. The paper has also been widely discussed at DANS. We would like to thank the following colleagues for their valuable comments, edits, pointers to documentation and discussions of the article set-up: Anja Smit, Ingrid Dillo, Hella Hollander, Valentijn Gilissen, Jan van Mansum, Ricarda Braukmann, Chris Baars, Kim Ferguson, Loek Brinkman and Nils Arlinghaus. This paper has been indirectly enabled by the following projects: FAIR Impact - Expanding FAIR Solutions across EOSC (EC ID 101057344); SSHOC-NL - Social Science and Humanities Open Science Cloud for the Netherlands (NWO 184.036.020); LTER-LIFE: a research infrastructure to develop Digital Twins of ecosystems in a changing world (NWO 184.036.014)

# Reference List

*About the umcs | NFU*. (2021, March 22). https://www.nfu.nl/en/nfu/about-umcs

Amsterdamska, O. (1987). Schools of Thought: Some Theoretical Observations. In O. Amsterdamska (Ed.), *Schools of Thought: The Development of Linguistics from Bopp to Saussure* (pp. 1–31). Springer Netherlands. https://doi.org/10.1007/978-94-009-3759-8_1

Borgman, C. L. (2010). *Scholarship in the digital age: Information, infrastructure, and the Internet*. MIT Press.

Borgman, C. L., & Groth, P. T. (2024). *From Data Creator to Data Reuser: Distance Matters* (No. arXiv:2402.07926). arXiv. https://doi.org/10.48550/arXiv.2402.07926

Borgman, C. L., Scharnhorst, A., & Golshan, M. S. (2019). Digital data archives as knowledge infrastructures: Mediating data sharing and reuse. *Journal of the Association for Information Science and Technology*, *70*(8), 888–904. https://doi.org/10.1002/asi.24172

Borgman, C. L., Van de Sompel, H., Scharnhorst, A., van den Berg, H., & Treloar, A. (2015). Who uses the digital data archive? An exploratory study of DANS. *Proceedings of the Association for Information Science and Technology*, *52*(1), 1–4. https://doi.org/10.1002/pra2.2015.145052010096

Borner, K. (2010). *Atlas of Science: Visualizing What We Know*. MIT Press.

Börner, K., Klavans, R., Patek, M., Zoss, A. M., Biberstine, J. R., Light, R. P., Larivière, V., & Boyack, K. W. (2012). Design and Update of a Classification System: The UCSD Map of Science. *PLOS ONE*, *7*(7), e39464. https://doi.org/10.1371/journal.pone.0039464

Braukmann, R. (2020). PID Graph van de Nederlandse wetenschap: DANS verbindt en verrijkt beschikbare informatie in NARCIS. *E-Data & Research*, *14*(3), 3.

Bryman, A. (2012). *Social Research Methods, 4th Edition* (4th edition). Oxford University Press.

Chang, H. (2016). *Autoethnography as Method*. Routledge. https://doi.org/10.4324/9781315433370

Christine L. Borgman. (2015). *Big Data, Little Data, No Data. Scholarship in the Networked World*. MIT Press. https://mitpress.mit.edu/big-data-little-data-no-data

Cook, T. (2011). The Archive(s) Is a Foreign Country: Historians, Archivists, and the Changing Archival Landscape. *The American Archivist*, *74*(2), 600–632. https://doi.org/10.17723/aarc.74.2.xm04573740262424

Cooper, D. M., & Springer, R. (2019). Data Communities. *Ithaka S+R*. https://sr.ithaka.org/publications/data-communities/

Cooper, R., & Lilyea, B. (2022). I'm Interested in Autoethnography, but How Do I Do It? *The Qualitative Report*. https://doi.org/10.46743/2160-3715/2022.5288

*DANS Self-Evaluation-Report*. (2024).

De Solla Price, D. J., & Beaver, D. (1966). Collaboration in an invisible college. *American Psychologist*, *21*(11), 1011–1018. https://doi.org/10.1037/h0024051

Dijk, E. M. S., Baars, C., Hogenaar, A. Th., & van Meel, M. (2006). NARCIS: The Gateway to Dutch Scientific Information. ELPUB 2006. In *Digital Spectrum: Integrating Technology and Culture* (pp. 49–58). ELPUB.

Doorn, P. K. (2020). Archiving and Managing Research Data: Data services to the domains of the humanities and social sciences and beyond: DANS in the Netherlands. *Der Archivar*, *73*(01), 44–50.

Doorn, P. K., Dillo, I., & Witkamp, P. (2014). Building a Federated Infrastructure for Preservation of and Access to Research Data in the Netherlands: The Front Office-Back Office Model. In D. Katre & D. Giaretta (Eds.), *APA/C-DAC International Conference on Digital Preservation and Development of Trusted Digital Repositories* (pp. 72–77). EXCEL INDIA PUBLISHERS.

Ellis, C., Adams, T. E., & Bochner, A. P. (2011). Autoethnography: An Overview. *Historical Social Research / Historische Sozialforschung*, *36*(4 (138)), 273–290.

Eronen, M. I., & Brooks, D. S. (2024). Levels of Organization in Biology. In E. N. Zalta & U. Nodelman (Eds.), *The Stanford Encyclopedia of Philosophy* (Summer 2024). Metaphysics Research Lab, Stanford University. https://plato.stanford.edu/archives/sum2024/entries/levels-org-biology/

*Focus on FAIR: DANS 2021-2025*. (2020).

Foo, Y. Y., Tan, K., Xin, X., Lim, W. S., Cheng, Q., Rao, J., & Tan, N. C. (2021). Institutional ethnography – a primer. *Singapore Medical Journal*, *62*(10), 507. https://doi.org/10.11622/smedj.2021199

GESIS. (1989). Steinmetz Archive: Dutch Social Science Data-Archive. *Historical Social Research / Historische Sozialforschung*, *14*(1 (49)), 118–121.

Glissen, V. (2014). *Dutch archaeological data depositing, processing, archiving and accessing at DANS: A repository with ten years of history, setting its sails to the future*.

Gregory, K., Groth, P., Scharnhorst, A., & Wyatt, S. (2020). Lost or Found? Discovering Data Needed for Research. *Harvard Data Science Review*, *2*(2). https://doi.org/10.1162/99608f92.e38165eb

Heidorn, P. B. (2008). Shedding Light on the Dark Data in the Long Tail of Science. *Library Trends*, *57*(2), 280–299.

Hof, C. H. J. (2019). *Beleidsnota DANS en de Levenswetenschappen* [Dataset]. DANS Data Station Life Sciences. https://doi.org/10.17026/dans-xqm-an3n

ISO. (2023). *ISO 14721:2012*. ISO. https://www.iso.org/standard/57284.html

Ketelaar, E. (1996). Archival theory and the Dutch Manual. *Archivaria*, 31–40.

Latour, B., & WooIgar, S. (1986). *Laboratory Life: The Construction of Scientific Facts*. Princeton University Press. https://doi.org/10.2307/j.ctt32bbxc

Lave, J., & Wenger, E. (1991). *Situated Learning: Legitimate Peripheral Participation*. Cambridge University Press.

Lavoie, B. F. (2004). The Open Archival Information System Reference Model: Introductory Guide. *Microform & Imaging Review*, *33*(2). https://doi.org/10.1515/MFIR.2004.68

Leonelli, S. (2023). *Philosophy of Open Science* (1st ed.). Cambridge University Press. https://doi.org/10.1017/9781009416368

Leonelli, S., & Ankeny, R. A. (2015). Repertoires: How to Transform a Project into a Research Community. *BioScience*, *65*(7), 701–708. https://doi.org/10.1093/biosci/biv061

Life sciences. (2024). In *Wikipedia* (2024th ed.). https://en.wikipedia.org/w/index.php?title=List_of_life_sciences&oldid=1257562118

Lorentz Center. (2021). *Center for Scientific Workshops in All Disciplines—FAIR Data for the 'Long Tail of Science.'* https://www.lorentzcenter.nl/fair-data-for-the-long-tail-of-science.html

*LTER-LIFE*. (2022, May 25). https://lter-life.nl/en

*Map of Science—UCSD map*. (2007). KNOWeSCAPE. http://knowescape.org/map-of-science-an-

update/50_______jpg_600x600_q85/

Marshall, C., & Rossman, G. B. (1989). *Designing Qualitative Research*. SAGE Publications.

Patton, M. Q. (2002). *Qualitative Research & Evaluation Methods*. SAGE.

PLAN-E. (2018). *Https://plan-europe.eu/wp-content/uploads/2018/10/report-plan-e-workshop-the-long-tail-of-science-and-data-version-1-0.pdf*. https://plan-europe.eu/wp-content/uploads/2018/10/report-plan-e-workshop-the-long-tail-of-science-and-data-version-1-0.pdf

Pratt, M. L. (1991). Arts of the Contact Zone. *Profession*, 33–40.

Rathenau Instituut. (2024). *Publication output by scientific field, international benchmark (WoS)*. https://www.rathenau.nl/en/science-figures/output/publications/publication-output-scientific-field-international-benchmark-wos

Rayward, W. B. (1991). The case of Paul Otlet, pioneer of information science, internationalist, visionary: Reflections on biography. *Journal of Librarianship and Information Science*, *23*(3), 135–145. https://doi.org/10.1177/096100069102300303

Scharnhorst, A., Bosch, O. ten, & Doorn, P. (2012). *Looking at a digital research data archive—Visual interfaces to EASY* (No. arXiv:1204.3200). arXiv. https://doi.org/10.48550/arXiv.1204.3200

Slavic, A. (2005). *Classification management and use in a networked environment: The case of the Universal Decimal Classification*. https://www.proquest.com/openview/8744572000fb8401422fc4a5267e7b3b/1?pq-origsite=gscholar&cbl=2026366&diss=y

Smith, D. E. (2005). *Institutional Ethnography: A Sociology for People*. Rowman Altamira.

*STZ*. (2024, December 13). https://www.stz.nl/. https://www.stz.nl/

Wallis, J. C., Rolando, E., & Borgman, C. L. (2013). If We Share Data, Will Anyone Use Them? Data Sharing and Reuse in the Long Tail of Science and Technology. *PLOS ONE*, *8*(7), e67332. https://doi.org/10.1371/journal.pone.0067332

Weinberg, A. M. (1961). Impact of Large-Scale Science on the United States. *Science*, *134*(3473), 161–

164. https://doi.org/10.1126/science.134.3473.161

Weisz, D. (1982). *Studies of Scientific Disciplines: An Annotated Bibliography*. Division of Planning and Policy Analysis, Office of Planning and Resources Management, National Science Foundation.

Wilkinson, M. D., Dumontier, M., Aalbersberg, Ij. J., Appleton, G., Axton, M., Baak, A., Blomberg, N., Boiten, J.-W., Santos, L. B. da S., Bourne, P. E., Bouwman, J., Brookes, A. J., Clark, T., Crosas, M., Dillo, I., Dumon, O., Edmunds, S., Evelo, C. T., Finkers, R., … Mons, B. (2016). The FAIR Guiding Principles for scientific data management and stewardship. *Scientific Data*, *3*, 160018. https://doi.org/10.1038/sdata.2016.18

Γληνός, K., & Glinos, K. (2010). *Riding the wave. How European can gain from the rising tide of the scientific data a vision for 2030*. https://lekythos.library.ucy.ac.cy/handle/10797/11910

# Figures and Tables

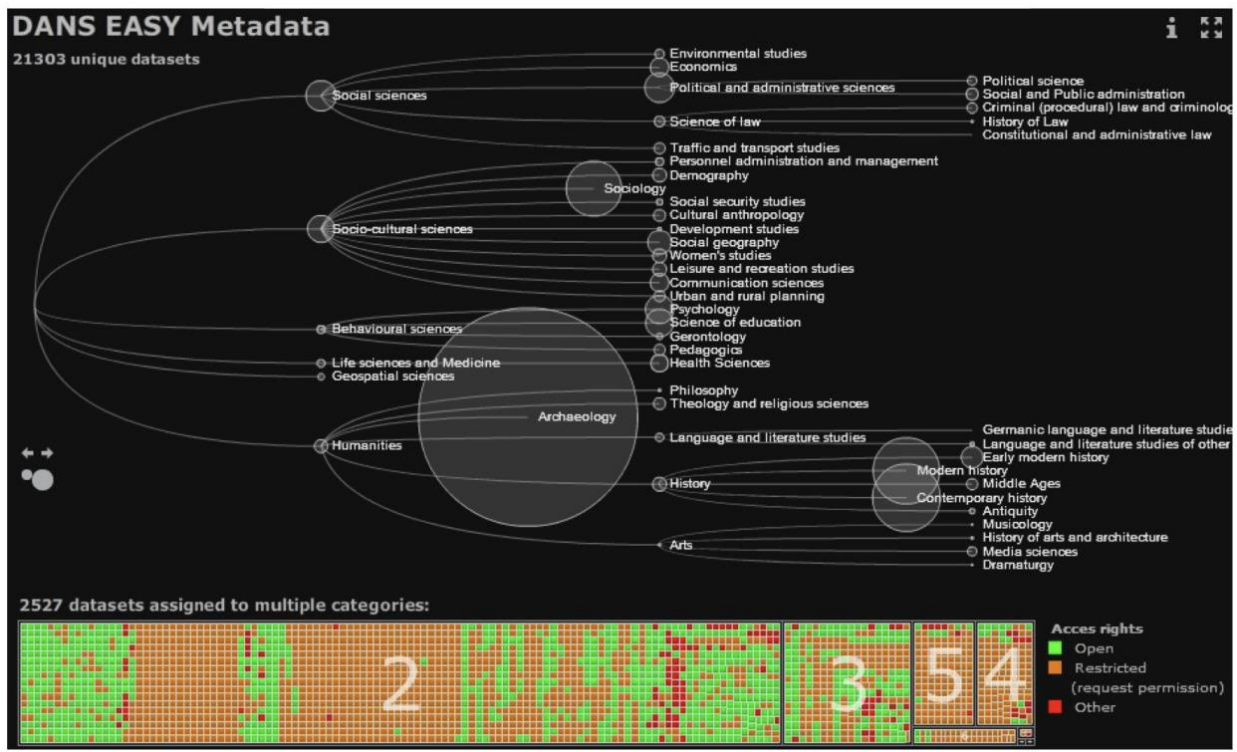

Fig 1: Analysis of the research fields used to index datasets in DANS-EASY (by 2012) - reproduction from (XXX et al., 2012).

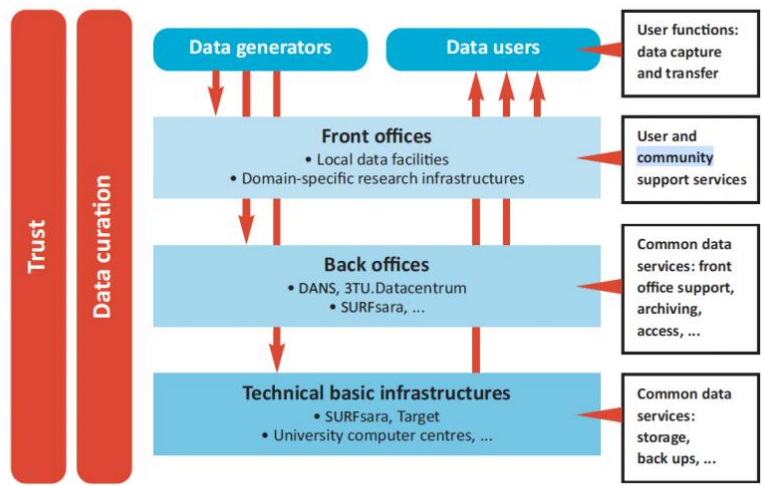

Fig. 2 Graph representing the back office/ front office infrastructure. DANS is present in the back office. Source: Internal Strategy Document (Doorn et al., 2014)

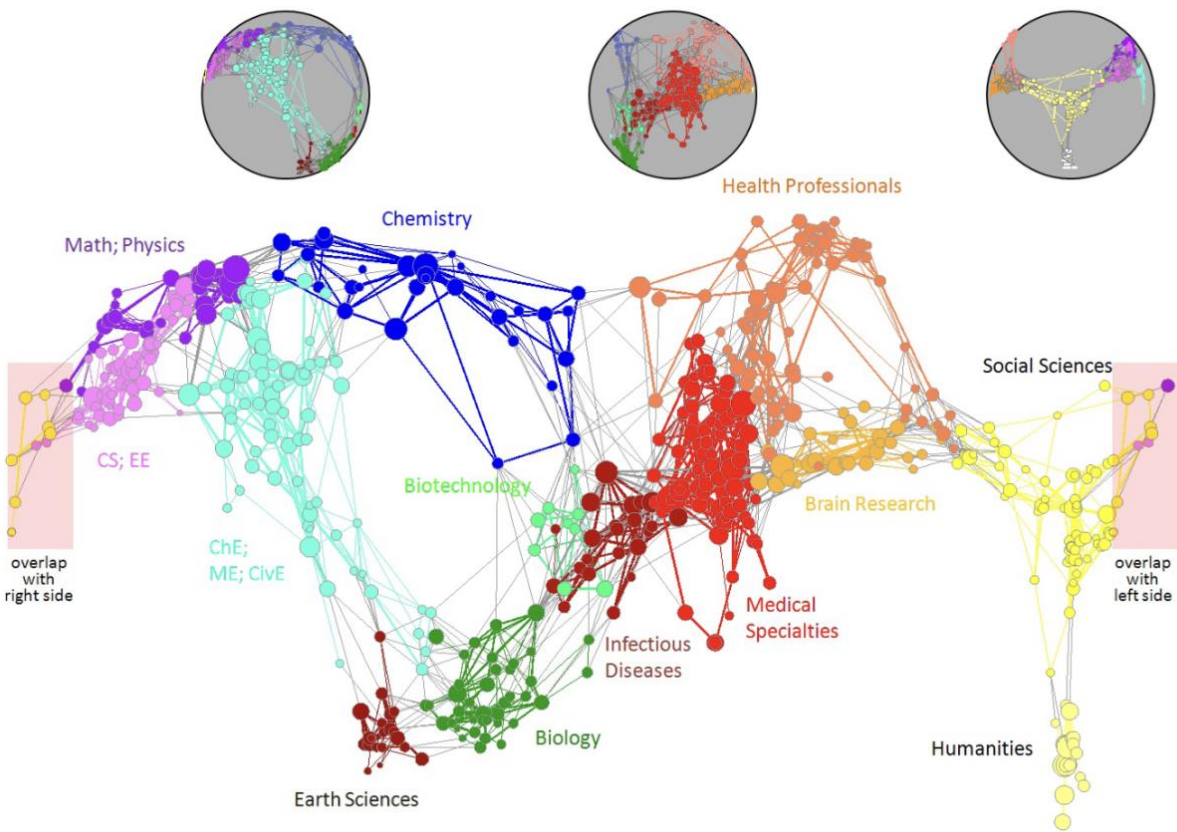

Figure 3: Part of the USCD Map